\documentclass[10pt,twocolumn,showpacs]{revtex4}
\usepackage{graphicx}
\usepackage{latexsym}
\usepackage{color}
\usepackage{amssymb}
\begin{document}
\title{One and two spin$-1/2$ particles systems under Lorentz
transformations}
\author{Hooman Moradpour$^1$\footnote{h.moradpour@riaam.ac.ir}, Mahdi Bahadoran$^2$\footnote{bahadoran@utm.my} and Ali Ghasemi$^1$}
\address{$^1$ Research Institute for Astronomy and Astrophysics of Maragha (RIAAM),
P.O. Box 55134-441, Maragha, Iran\\
$^2$ Laser Centre, Ibnu Sina, Institute for Scientific and
Industrial Research (ISI-SIR) Universiti Teknologi Malaysia (UTM),
81310, Johor Bahru, Malaysia}
\begin{abstract}
Lorentz transformation (LT) is used to connect two inertia frames,
including the lab and moving frames, and the effect of LT on the
states of one spin-1/2 particle system is studied. Moreover, we
address the predictions made by Czachor's and the Pauli spin
operators about the spin behavior and compare our results with the
behavior of system's state under Lorentz transformation. This
investigation shows that the predictions made by considering the
Pauli spin operator about the spin of system are in better
agreement with the system's state in comparison with that of made
by considering Czachor's spin operator. In continue, we focus on
two-particle pure entangled systems including two spin$-1/2$
particles which moves away from each other. Once again, the
behavior of this system's states under Lorentz transformation are
investigated. We also point to the behavior of Bell's inequality,
as a witness for non-locality, under Lorentz transformation. Our
study shows that the Bell operator made by the Pauli operator has
better consistency with the behavior of spin state of system under
Lorentz transformation compared with the Bell operator made by
Czachor's operator. Our approach can be used to study the relation
between the various spin operators and the effect of LT on system,
which provides the basement to predict the outcome of a
Stern-Gerlach type experiment in the relativistic situations.
\end{abstract}
\pacs{03.30.+p, 03.65.Yz, 03.65.Ud}
\maketitle
{Keywords: Spin; lorentz transformation; non-locality}

\section{Introduction \label{Introduction}}
In quantum mechanics, systems may blurt a non-local behavior from
themselves \cite{EPR}. Bohm and Aharanov provided a spin version
for exhibiting this behavior \cite{BA}. In their setup,
non-locality leads to entanglement, i.e. the state of system is
not equal to the product of its constituent particles' states
\cite{BA}. Firstly, Bell tried to get a criterion for
distinguishing the local and non-local phenomenons from each other
\cite{Bell}. His work leads to a well-known inequality called the
Bell inequality which may be violated by non-local states. In
fact, There are various models for this inequality
\cite{CHSH,aud,rev,Bert}. In the two-particle systems, the Bell
operator is defined as \cite{CHSH}
\begin{eqnarray}\label{bel}
B=a\otimes (b+b') + a'\otimes(b-b'),
\end{eqnarray}
where ($a,a'$) and ($b,b'$) are yes or no operators applying on
the first and second particles, respectively. For every local
state, the Bell operator meets the $\langle B \rangle\leq 2$
condition \cite{CHSH}.

Some forehand experimental attempts have been done to detect
non-locality can be found in \cite{aspect1,aspect2,aspect3}. It is
shown that non-locality is not limited to the multi-particle
systems and indeed, a one-particle system may also behave
non-locally \cite{vedral1,vedral2}. Non-locality is a source for
entropy which has vast implications in current science
\cite{aud,Nilsen}. It has also been shown that it may be a source
for the entropy of horizons in the gravitational and cosmological
setups \cite{grav1}.

Spin is a quantum mechanical property of systems which was exhibited
in investigating the relativistic quantum mechanical systems. Pauli
derived an operator for describing the spin of particles in the low
velocity limit. By considering the low velocity limit, Pauli got
$2\times2$ matrixes, called the Pauli matrixes or operator
($\sigma_i$), and the corresponding spin operator
$S_i=\frac{\hbar}{2}\sigma_i$, $\forall i=x,y,z$ for spin-$1/2$
particles \cite{grain}. Nowadays, it is believed that the
predictions made by the Pauli spin operator ($S_i$) about the spin
of systems are in line with the Stern-Gerlach type experiments in
the lab frame \cite{sakur}. But, is it the only candidate for the
spin operator which leads to the consistent results with a
Stern-Gerlach type experiment in the lab frame? Moreover, what is
the result of a Stern-Gerlach type experiment, if it is observed by
a moving observer which moves with respect to the lab frame with a
constant velocity ($\beta$)? Indeed, there are various attempts to
get a candidate for describing spin and thus the results of applying
a Stern-Gerlach type experiment on a system which is in relative
motion with respect to observer
\cite{czachor,spin1,spin2,spin4,caban2012,caban2013}.

Czachor followed the Pryce \cite{pry} and Fleming \cite{fle}
arguments to get
\begin{eqnarray}\label{cza}
\widehat{A}=\frac{(\sqrt{1-\beta^2}\overrightarrow{A}_{\bot}
+\overrightarrow{A}_{\|}).\overrightarrow{\sigma}}{\sqrt{1+\beta^2[(\widehat{e}
.\overrightarrow{A})^2-1]}},
\end{eqnarray}
as the spin operator along the unit vector $\overrightarrow{A}$,
which commutes with the Hamiltonian \cite{czachor}. Based on this
result, this operator may be used instead of the Pauli operator
along the $\overrightarrow{A}$ vector
($\overrightarrow{A}.\overrightarrow{\sigma}$) whenever, states
with zero momentum uncertainty are taken into account
\cite{czachor}. Here, $\sigma$ and $\widehat{e}$ are the Pauli
operator and the unit vector along the boost direction,
respectively. Additionally, $\beta$ is the boost velocity, and,
independent of $\overrightarrow{A}$, the Pauli spin operator is
recovered by substituting $\beta=0$. Moreover, the subscripts
$\bot$ and $\|$ denote the perpendicular and parallel components
of the vector $\overrightarrow{A}$ to the boost direction. This
operator also covers the Pauli spin operator whenever either
$\overrightarrow{A}_{\|}=0$ or $\overrightarrow{A}_{\bot}=0$
meaning that $\widehat{A}=\overrightarrow{A}.\widehat{\sigma}$. It
is worth to note that the uncertainty principle leads to
$\triangle\beta\neq0$ and therefore, this principle prevents such
possibility in a realistic experiment \cite{czachor}. Its
generalization to the wave-packets can be found in \cite{peres5}.
Some of the shortcomings and strengths of Czachor's and the Pauli
operators are investigated in Refs.
\cite{czachor,spin1,spin2,spin4}. Although, just the same as the
Pauli operator, Czachor's spin operator should indeed be defined
as $C=\frac{\hbar}{2}\widehat{A}$ to coverer the spin-$1/2$
particles, but, we should note that the eigenvalues of Czachor's
spin operator are not always equal to $\pm\frac{\hbar}{2}$
\cite{spin1,spin2}.

Whenever the effects of considering high velocities such as the
probability of pair production are ignored, the quantum mechanical
interpretations of phenomena are satisfactory and the lab frame is
connected to the moving frame, which moves with a constant velocity
with respect to the lab frame, by a LT \cite{hal}. Therefore, one
may apply LT on the system state in the lab frame to get state seen
by the moving observer. By this approach, the spin state of system
is affected by a rotation with the Wigner angle \cite{wigner}. The
effects of LT on the single-particle entangled states are
investigated by Palge et al. \cite{pvedral}. It is shown that such
rotations may also affect the spin entropy of one spin $1/2$
particle as well as the two spin $1/2$ entangled particles systems
\cite{pvedral1,peres1,IJMPA}. There are also various attempts in
which authors investigate the behavior of non-locality under LT
\cite{gingrich,li,jordan,alsing,terashima,terashima1,ahn,lee,kim,caban2009,saldanha,saldanha1,friis,AA,you,moradi1,moradi2,moradi3,moradi4,mmm,mm}.
Their results can also be used to get some theoretical predictions
about the outcome of a Stern-Gerlach type experiment which may lead
to get a more suitable spin operator. The acceleration effects on
non-locality are also investigated in
\cite{fu,al,ma,le,fun,smith,alsing,tera,shi,ball,ver}.

Some authors have used the Pauli spin operator to build the Bell
operator and considered bi-partite pure entangled state
\cite{terashima,terashima1}. Thereinafter, they considered a
special set of measurement directions which leads to violate
Bell's inequality to its maximum violation amount ($2\sqrt{2}$) in
the lab frame. In addition, they have been considered a moving
observer connected to the lab frame by a LT, and applied LT on the
system state in the lab frame to get the corresponding state in
the moving frame. They took into account the same set of
measurement directions for the moving frame as the lab frame, and
investigate the behavior of Bell's inequality in the moving frame.
In fact, they use Bell's inequality as a witness for the
bi-partite non-locality. Finally, they find that the violation of
Bell's inequality in the moving frame is decreased as a function
of the boost velocity and the particles energy in the lab frame
\cite{terashima,terashima1}. It should be noted that if one
applies LT on both of the Bell operator and the system state,
Bell's inequality is violated to the same value as the lab frame
\cite{terashima,terashima1}. The generalization of this work to
three-particle non-local systems can be found in \cite{mmm,you}.

In a similar approach, Ahn et al. have been considered the Bell
states and used Czachor's operator to construct the Bell operator
\cite{ahn}. Bearing in mind this fact that Czachor's and the Pauli
operators are the same operators in the lab frame ($\beta=0$),
authors have considered the special set of spin measurements which
violates Bell's inequality to its maximum violation amount in the
lab frame. They applied LT on the system state in the lab frame to
get the corresponding state in the moving frame. They also assumed
that the moving frame uses the same set of spin measurements as the
lab frame for evaluating Bell's inequality. Therefore, their setup
has some similarity with those of Refs.~\cite{terashima,terashima1}.
There are also some differences between setups investigated in these
papers. Their LT differs from each other, and they used different
spin operator to build the Bell operator. Finally, Ahn et al. found
out that the expectation value of the Bell operator in the moving
frame is decreased as a function of the boost velocity and the
energy of particles in the lab frame \cite{ahn}. It should be noted
again that Bell's inequality will be violated in the moving frame to
the same value as the lab frame, if one applies LT on both of the
Bell operator and the system state \cite{kim,lee,friis}. More
studies on this subject and its generalization to the three-particle
non-local systems can be found in
\cite{saldanha,saldanha1,friis,moradi4,moradi1,moradi2,moradi3,mm}.

In fact, both of the mentioned approaches found out that the
expectation value of the Bell operator in the moving frame is
decreased by increasing the boost velocity and the energy of
particles in the lab frame. Although it seems that this conclusion
is a common result between the mentioned attempts, but they are
completely different from each other. For example, based on the
results obtained in \cite{terashima,terashima1}, Bell's
inequality, in the moving frame and the $\beta\rightarrow1$ limit,
is violated to its maximum violation amount for the low energy
particles, whilst the results observed by Ahn et al. suggest that
this inequality is preserved at this limit independent of the
particles energy. However, the question is which view is correct?
Here we used the Stern-Gerlach type experiment as an appropriate
approach to solve this problem. Is it possible to get more
theoretical information about this inconsistency appeared in these
studies? Indeed, this inconsistency between the results of
considering Czachor's operator and those of considered the Pauli
operator will be more complicated in the three-particle non-local
systems \cite{mmm}. It is reported that the Pauli operator and its
corresponding spin operator lead to better agreement with the
behavior of spin state of the three-particle non-local systems
under LT compared with Czachor's operator \cite{mmm}. Moreover,
the Pauli-Lubanski spin operator is not suitable to describe the
spin interaction with a magnetic field in the moving frame
connected to the lab frame by a LT \cite{njp}. Indeed, authors
took into account the Pauli-Lubanski definition of spin operator
and introduced a Hamiltonian for spin interaction with magnetic
field. In continue, they considered the effects of LT on the
reduced spin density matrix of one spin-$\frac{1}{2}$ particle and
the results of applying a Stern-Gerlach experiment on the system
in various frames by focusing on the quantization axes in the
various frames \cite{njp}. Finally, they concluded that the
Pauli-Lubanski spin operator (and similar operators such as
Czachor's operator) is not suitable for describing the system,
which includes a one spin-$\frac{1}{2}$ particle interacting with
magnetic field in the lab frame, in all inertial frames connected
to each other by LT \cite{njp}. Hence, what is the origin of these
differences between the results of considering the Pauli operator
and that of Czachor? Loosely speaking, which one of these
operators is in better agreement with the spin states of one and
two-particle systems, and helps us get more suitable predictions
about the results of applying a Stern-Gerlach type experiment on a
system which is in relative motion with respect to observer?

In this paper, we study the differences between the results of
considering the Pauli operator for describing spin and those of used
Czachor's operator to investigate spin. Unlike Ref.~\cite{njp}, we
do not consider any magnetic field. Moreover, in order to avoid any
paradoxes due to apply LT on the system \cite{saldanha,saldanha1},
we consider a situation in which the particles momentums are
specified with zero uncertainty, in the lab frame. In addition, we
focus on the behavior of the system state and spin operator, and
show that, even in the absence of the magnetic fields, Czachor's
spin operator is not probably suitable to describe spin. We start
from the one particle system and consider a moving observer
connected to the lab frame by a LT. By studying the behavior of spin
state in the lab and moving frames, we try to establish a
theoretical criterion to decide about the validity of spin behavior
predicted by either using Czachor's or the Pauli spin operators. In
addition, we generalize our study to the two-particle non-local
system. Our results indicate that the Pauli operator is in better
agreement with the behavior of spin state in both of the lab and
moving frames compared with Czachor's operator.

The paper is organized as follows. In the next section, we focus on
the one particle system and investigate the behavior of expectation
values of Czachor's and the Pauli spin operators under LT. We also
point to the behavior of the spin state under LT, and compared the
results with the behavior of expectation values of Czachor's and the
Pauli spin operators under LT to get the better spin operator. In
section ($\textmd{III}$), we focus on the two-particle non-local
system including two purely entangled spin$-1/2$ particles which
moves away from each other along the $z$ direction with the same
momentum. In addition, we point to the above mentioned inconsistency
and try to eliminate that by considering the behavior of the system
state under LT. The last section is devoted to the summary and
concluding remarks. Throughout this paper we set $c=1$ for
simplicity.
\section{Quantum Mechanics under LT}
In the lab frame ($S$) for a spin-$\frac{1}{2}$ particle, with the
momentum state $\vert \overrightarrow{p}\rangle$ and the spin
state $\vert\Sigma\rangle$, the state of system is written as
\begin{eqnarray}
\vert \xi\rangle=\vert
\overrightarrow{p}\rangle\vert\Sigma\rangle.
\end{eqnarray}
Here, we take into account that $\overrightarrow{p}=p_0 \hat{z}$.
The state of particle is viewed by an observer which moves along
the $x$ axis $(\overrightarrow{\beta}=\beta\widehat{x})$ as
\begin{eqnarray}
\vert \xi \rangle^{\Lambda}=\vert
\overrightarrow{p}\rangle^{\Lambda} D(W(\Lambda,p_1))
\vert\Sigma\rangle,
\end{eqnarray}
where $\vert \overrightarrow{p}\rangle^{\Lambda}$ denotes the
momentum state of particle in the moving frame ($S'$), and
$D(W(\Lambda,p))$ is the spin-$\frac{1}{2}$ Wigner representation
of the Lorentz group \cite{wigner,hal}:
\begin{eqnarray}\label{wr}
D(W(\Lambda,p))=\cos\frac{\Omega_{p}}{2}-i\sigma_y
\sin\frac{\Omega_{p}}{2}.
\end{eqnarray}
In this equation, $\sigma_y$ and $\Omega_{p}$ are the Pauli matrix
and the Wigner angle, respectively, evaluated as
\begin{eqnarray}\label{angle}
\tan\Omega_p=\frac{\sinh\alpha\sinh\delta}{\cosh\alpha +
\cosh\delta}.
\end{eqnarray}
$\cosh\delta=\frac{p_0}{m}$ and $\cosh\alpha=\sqrt{1-\beta^2}$ are
related to the particle energy in the lab frame and the boost
effects, respectively. Therefore, simple calculations lead to
\begin{eqnarray}\label{state}
\vert + \rangle^{\Lambda}&=&\cos\frac{\Omega_{p}}{2}\vert +
\rangle
+ \sin\frac{\Omega_{p}}{2}\vert - \rangle, \nonumber \\
\vert - \rangle^{\Lambda}&=&\cos\frac{\Omega_{p}}{2}\vert -
\rangle - \sin\frac{\Omega_{p}}{2}\vert +\rangle,
\end{eqnarray}
where $\vert + \rangle$ and $\vert - \rangle$ denote the up and
down spin states along the $z$ direction in the lab frame,
respectively. Moreover, the superscript $\Lambda$ is used to
specify the corresponding spin state in the moving frame. In the
$\beta\rightarrow1$ limit,
$\sin\frac{\Omega_p}{2}\sim\sqrt{\frac{\Gamma-1}{2\Gamma}}$ where
$\Gamma=\frac{1}{\sqrt{1-\beta_1^2}}$ and $\beta_1$ are the energy factor
and the velocity of particle in the lab frame, respectively. Using
Eq.~(\ref{state}) to get
\begin{eqnarray}\label{state1}
\vert + \rangle^{\Lambda}& \approx & \vert + \rangle, \nonumber \\
\vert - \rangle^{\Lambda}& \approx &\vert - \rangle,
\end{eqnarray}
for a low energy particle ($\Gamma\rightarrow 1$) and
\begin{eqnarray}\label{state2}
\vert + \rangle^{\Lambda}& \approx &\frac{1}{\sqrt2}( \vert + \rangle + \vert - \rangle), \nonumber \\
\vert - \rangle^{\Lambda}& \approx &\frac{1}{\sqrt2}( \vert
-\rangle - \vert + \rangle),
\end{eqnarray}
for a high energy particle ($\Gamma\rightarrow \infty$). Now,
consider a situation in which the lab and moving observers apply a
Stern-Gerlach experiment in the same direction
$\overrightarrow{A}=(\frac{1}{\sqrt{2}},0,\frac{1}{\sqrt{2}})$,
while, $\vert + \rangle$ is the spin state of particle in the lab
frame. Therefore, the spin state of system in the moving frame can
be found in Eq.~(\ref{state}). Let us focus on the results
obtained by taking into account Czachor's and the Pauli spin
operators. If the lab and moving frames use the Pauli spin
operator and the same spin measurement direction
($\overrightarrow{A}$), simple calculations lead to
\begin{eqnarray}\label{pauli10}
\langle S \rangle_1= \frac{\hbar}{2\sqrt{2}},
\end{eqnarray}
and
\begin{eqnarray}\label{pauli1}
\langle S \rangle_2= \frac{\hbar}{2\sqrt{2}}(\cos\Omega_p
+\sin\Omega_p),
\end{eqnarray}
for the lab and moving frames, respectively. Here, the subscript $1$ and $2$ are also used to denote the lab and moving frames, respectively. In addition, $S=\frac{\hbar}{2}(\overrightarrow{A}.\overrightarrow{\sigma})=\frac{\hbar}{2\sqrt{2}}(\sigma_x
+\sigma_z)$ is the spin operator along the $\overrightarrow{A}$
direction. It is easy to check that Eq.~(\ref{pauli1}) is in line
with the asymptotic behavior explained in Eqs.~(\ref{state1})
and~(\ref{state2}). It is useful to mention here that, in the
$\beta\rightarrow1$ limit, Eq.~(\ref{pauli1}) leads to
$\frac{\hbar}{2\sqrt{2}}$ for both of the low ($\Omega_p\sim0$)
and high ($\Omega_p\sim\frac{\pi}{2}$) energy particles. This
result is in agreement with the asymptotic behaviors addressed in
Eqs.~(\ref{state1}) and~(\ref{state2}).

If Czachor's spin operator together with the vector
$\overrightarrow{A}=(\frac{1}{\sqrt{2}},0,\frac{1}{\sqrt{2}})$ are
considered, one can use Eq.~(\ref{cza}) in order to evaluate the
spin operator in the lab and moving frame as
\begin{eqnarray}\label{cza1}
C_1=\frac{\hbar}{2\sqrt{2}}(\sigma_x +\sigma_z),
\end{eqnarray}
and
\begin{eqnarray}\label{cza2}
C_2=\frac{\hbar}{2}(\frac{\sigma_x
+\sqrt{1-\beta^2}\sigma_z}{\sqrt{2-\beta^2}}),
\end{eqnarray}
respectively. In this equations, the subscripts $1$ and $2$ denote
the lab and moving frames, respectively. Since $\vert + \rangle$
is the spin state of particle in the lab frame, by using
Eq.~(\ref{cza1}) we get
\begin{eqnarray}\label{cza10}
\langle C_1 \rangle= \frac{\hbar}{2\sqrt{2}},
\end{eqnarray}
which is the same as the previous results obtained by considering
the Pauli spin operator. For the moving frame, using
Eqs.~(\ref{state}) and~(\ref{cza2}) to obtain
\begin{eqnarray}\label{cza010}
\langle C_2 \rangle=
\frac{\hbar}{2\sqrt{2-\beta^2}}(\sqrt{1-\beta^2}\cos\Omega_p+\sin\Omega_p),
\end{eqnarray}
which differs from the result obtained by using the Pauli
operator, Eq.~(\ref{pauli1}). But which approach is right? In
order to check this prediction, we focus on the
$\beta\rightarrow1$ limit. For the low energy particle
$\Omega_p\rightarrow0$ and we get $\langle C_2 \rangle
\rightarrow0$ which is fully inconsistent with Eq.~(\ref{state1}).
Additionally, considering $\Omega_p\rightarrow\frac{\pi}{2}$ leads
to $\langle C_2 \rangle \rightarrow\frac{\hbar}{2}$ which is again
in contrast with the result predicted by Eq.~(\ref{pauli1}). All
in all, using both of these operators lead to the same predictions
for the spin in the lab frame but, in the moving frame the results
obtained using Czachor's spin operator differ from that of
considered the Pauli spin operator. Our approach shows that the
predictions made by considering the Pauli spin operator is in
agreement with the behavior of the spin state in the lab and
moving frames. The same conclusion is not accessible by
considering Czachor's spin operator which indicates that our
approach is in full agreement with previous results obtained in
\cite{njp}. We should note that the Stern-Gerlach type experiment
is required to identify the correct result between
Eq.~(\ref{pauli1}) or Eq.~(\ref{cza010}). Based on
Eq.~(\ref{cza}), in the moving frame,
$C=\frac{\hbar}{2}\widehat{A}=\frac{\hbar}{2}(\overrightarrow{A}.\widehat{\sigma})=\frac{\hbar}{2}S$
under one of the the following conditions
$\widehat{e}.\overrightarrow{A}=0$ ($\overrightarrow{A}_{\|}=0$)
or $\widehat{e}.\overrightarrow{A}=1$
($\overrightarrow{A}_{\perp}=0$). Under this condition Czachor's
operator is the same as the Pauli operator and the results lead to
the same predictions in the lab and moving frames.
\section{Pure bi-partite entangled states under LT}
In order to investigate the LT effects on the pure bi-partite
non-locality, the above arguments is required to be generalized to
the two-particle system. This generalization is as follows. For a
system, including two spin-$\frac{1}{2}$ particles, in the lab
frame ($S$), with the spin state $\vert\psi\rangle$ and the
momentum state $\vert
\overrightarrow{p_1}\overrightarrow{p_2}\rangle$, the state of
system is
\begin{eqnarray}
\vert \xi\rangle=\vert
\overrightarrow{p_1}\overrightarrow{p_2}\rangle\vert\psi\rangle.
\end{eqnarray}
Now, consider a moving frame ($S'$) which moves along the $x$ axis
$(\overrightarrow{\beta}=\beta\widehat{x})$. In the $S'$ frame,
the state of system is
\begin{eqnarray}
\vert \xi \rangle^{\Lambda}=\vert
\overrightarrow{p_1}\overrightarrow{p_2} \rangle^{\Lambda}
\prod_{i=1}^{2} D(W(\Lambda,p_i)) \vert\psi\rangle.
\end{eqnarray}
$\vert \overrightarrow{p_1}\overrightarrow{p_2}\rangle^{\Lambda}$
denotes the momentum state of system in the moving frame, and
$D(W(\Lambda,p_i))$ is again the spin-$\frac{1}{2}$ Wigner
representation of the Lorentz group for the i$^{\textmd{th}}$
particle Eq.~(\ref{wr}). Consider a system, including two
particles which moves away from each other along the $z$ direction
in the lab frame, with the total state as
\begin{eqnarray}\label{slab}
\vert \xi \rangle=\vert
\overrightarrow{p_1}\overrightarrow{p_2}\rangle\frac{1}{\sqrt{2}}(\vert
++\rangle + \vert - -\rangle ),
\end{eqnarray}
where $\overrightarrow{p_1}=-\overrightarrow{p_2}=p\widehat{z}$.
If the moving frame considers the same basis as the lab frame, and
applying LT on the system state, the system state for the moving
frame is given by \cite{ahn,kim}
\begin{eqnarray}\label{slab1}
\vert \xi \rangle^{\Lambda}&=&\frac{\vert
\overrightarrow{p_1}\overrightarrow{p_2}\rangle^{\Lambda}}{\sqrt{2}}(\cos\Omega_p(\vert
++\rangle + \vert - -\rangle)\nonumber \\
&-&\sin\Omega_p(\vert +-\rangle - \vert
- +\rangle)).
\end{eqnarray}
The maximum violation of Bell's inequality in the lab frame
($\langle B \rangle=2\sqrt{2}$) is obtainable by choosing the
\begin{eqnarray}\label{bel1}
\overrightarrow{a}=(\frac{1}{\sqrt{2}},-\frac{1}{\sqrt{2}},0),\ \
\overrightarrow{a}'=(-\frac{1}{\sqrt{2}},-\frac{1}{\sqrt{2}},0),
\end{eqnarray}
and
\begin{eqnarray}\label{bel2}
\overrightarrow{b}=(0,1,0),\ \ \overrightarrow{b}'=(1,0,0),
\end{eqnarray}
for the directions of Pauli's operators applying on the first
and second particles, respectively \cite{aud,ahn,kim}. As noted in
the introduction, since $\beta$ meets the $\beta=0$ condition in
the lab frame, Czachor's operator is compatible with the Pauli
operator in the lab frame. Therefore, this result is also
obtainable in the lab frame if Czachor's operator is used to build
the Bell operator. Let us focus on the moving observer. If the
moving observer uses Czachor's operator and the special set of
measurement directions, given in Eqs.~(\ref{bel1})
and~(\ref{bel2}), then \cite{ahn}
\begin{eqnarray}\label{cza3}
\langle B_C
\rangle=\frac{2}{\sqrt{2-\beta^2}}(\sqrt{1-\beta^2}+\cos\Omega_p),
\end{eqnarray}
here C index indicates the consideration of Czachor's operator.
The result of lab frame is obtainable by inserting $\beta=0$ and
$\Omega_p=0$ simultaneously. It is obvious that $\langle B_C
\rangle\leq2$ for the $\beta\rightarrow1$ limit,which means that
Bell's inequality is preserved in the moving frame and is
independent from the particles energy in the lab frame. This
behavior indicates that non-locality is vanished in this limit if
the moving frame uses the same set of measurements as the lab
frame violating the Bell inequality to its maximum violation
amount in the lab frame \cite{ahn}. From Eq.~(\ref{slab1}), we see
that, in the $\beta\rightarrow1$ limit and for the low energy
particles system ($\Omega_p\sim0$), $\vert \xi
\rangle^{\Lambda}\simeq\vert
\overrightarrow{p_1}\overrightarrow{p_2}\rangle^{\Lambda}\frac{1}{\sqrt{2}}(\vert
++\rangle + \vert - -\rangle)$ which is the same as the system
state in the lab frame. Therefore, the bi-partite non-locality
does not completely disappear at this limit. We believe a true
Bell operator should show this behavior. Therefore, we see that,
once again, Czachor's operator does not lead to the results
compatible with the behavior of the spin state of system. This
weakness of Czachor's operator was also reported in the
multi-particle non-local systems \cite{mmm}. Using the Pauli
operators to form the Bell operator in the moving frame, and
applying that on the spin state of system, we get
\begin{eqnarray}
\langle B_l \rangle = 2\sqrt{2}\cos^2\Omega_p,
\end{eqnarray}
here $l$ index indicates that the Pauli operators are used to
build the Bell operator. Moreover, we took into account the
special directions explained in Eqs.~(\ref{bel1})
and~(\ref{bel2}). $\langle B_l \rangle$ is displayed in Fig.~($1$).
\begin{figure}[ht]
\centering
\includegraphics[scale=0.4]{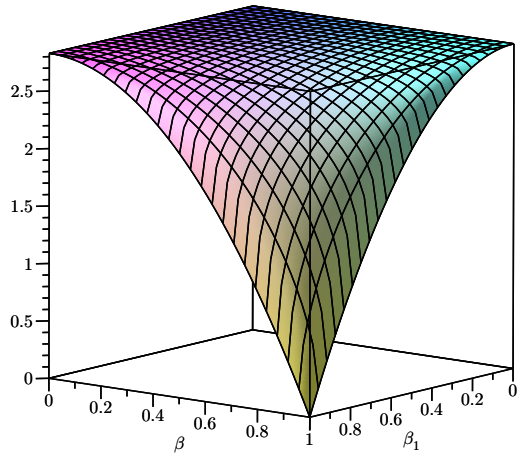}
\caption{The plot depicts $\langle B_l \rangle$. Here, $\beta$ and $\beta_1$ point to the boost velocity and the velocity of particles in the lab frame.}
\end{figure}
In the non-relativistic limit
($\Omega_p\rightarrow0$), the result of lab frame is obtainable.
In addition, for the low energy particles in the
$\beta\rightarrow1$ limit, $\Omega_p\rightarrow0$ and therefore,
Bell's inequality is maximally violated which is in agreement with
the asymptotic behavior of the system state. For the high energy
particles in the $\beta\rightarrow1$ limit,
$\Omega_p\rightarrow\frac{\pi}{2}$ which leads to $\vert \xi
\rangle^{\Lambda}\simeq\vert
\overrightarrow{p_1}\overrightarrow{p_2}\rangle^{\Lambda}\frac{1}{\sqrt{2}}(\vert
-+\rangle - \vert + - \rangle)$. The latter points that since the
spin measurement directions have not been changed in our setup,
Bell's inequality should be satisfied in the $\beta\rightarrow1$
limit by the high energy particles system. It is easy to check
that this expectation is satisfied by both of $\langle B_C
\rangle$ and $\langle B_l \rangle$. Finally, we found that,
independent of using either Czachor's or the Pauli spin operators
to describe the spin, the expectation amount of the Bell operator
is decreased as a function of the boost velocity and the particles
energy if the moving and lab frames use the same set of
measurement directions for the Bell operator violating Bell's
inequality to its maximum violation amount in the lab frame.

Another special case is
\begin{eqnarray}\label{tlab0}
\vert \xi \rangle=\vert
\overrightarrow{p_1}\overrightarrow{p_2}\rangle\frac{1}{\sqrt{2}}(\vert
+-\rangle + \vert - +\rangle ),
\end{eqnarray}
where again
$\overrightarrow{p_1}=-\overrightarrow{p_2}=p\widehat{z}$. This
state violates Bell's inequality to its maximum violation amount
($2\sqrt{2}$) in the lab frame by choosing
$\overrightarrow{a}=(\frac{1}{\sqrt{2}},\frac{1}{\sqrt{2}},0)$,
$\overrightarrow{a}'=(\frac{-1}{\sqrt{2}},\frac{-1}{\sqrt{2}},0),$
$\overrightarrow{b}'=(1,0,0)$ and $\overrightarrow{b}'=(0,1,0)$
\cite{aud,ahn}. In the moving frame, this state is given as
\cite{ahn}
\begin{eqnarray}\label{tlab}
\vert \xi \rangle^{\Lambda}=\vert
\overrightarrow{p_1}\overrightarrow{p_2}\rangle^{\Lambda}\frac{1}{\sqrt{2}}(\vert
+-\rangle + \vert - +\rangle ),
\end{eqnarray}
which means that the considered LT leaves this state unchanged.
Therefore, it is crystal clear that when the moving observer uses
the Pauli spin operator and the same set of measurements as the
lab frame to get the Bell operator and Bell's inequality, Bell's
inequality is violated to the same value as the lab frame. The
latter means that, in the moving frame, this inequality is
violated to its maximum violation amount ($2\sqrt{2}$) in this
situation. If the moving observer uses Czachor's operator to
construct the Bell operator and investigates the behavior of
Bell's inequality and thus the corresponding non-locality, then
\cite{ahn}
\begin{eqnarray}\label{cza34}
\langle B_C
\rangle=\frac{2}{\sqrt{2-\beta^2}}(\sqrt{1-\beta^2}+1),
\end{eqnarray}
which claims that non-locality is decreased by increasing the
boost velocity and thus, in the $\beta\rightarrow1$ limit, Bell's
inequality is marginally satisfied in the moving frame \cite{ahn}.
This result is in contrast with the invariant form of this state
under LT Eq.~(\ref{tlab}) and the results made by considering the
Pauli operator. Once again, It is figured out that, whenever the
lab and moving frames use the same set of measurements which
violate Bell's inequality to its maximum violation amount in the
lab frame, the behavior of the bi-partite pure entangled
state~(\ref{tlab}) under LT is fully consistent with the behavior
of the Bell operator under LT if the Pauli operator applied to
form the Bell operator. Loosely speaking, the same as the results
of the previous section and the three-particle non-local systems
\cite{mmm}, the predictions of Czachor's spin operator about the
spin differ from those of the Pauli operator and the system state.
It is useful to note here that a Stern-Gerlach type experiment is
needed to experimentally distinguish these results. Finally, we
should note that since LT introduced in this paper is a unitary
operator, it should be possible to get the same violation amount
as the lab frame for Bell's inequality in the moving frame.
Indeed, if one applies LT on both of the system state and the Bell
operator, Bell's inequality is also maximally violated in the
moving frame \cite{terashima,terashima1,kim,mmm}.
\section{Summary and Conclusion \label{Summary}}
Spin is a quantum mechanical property of systems. It has vast
implications in the spectroscopy, quantum information theory and
etc. Therefore, it is necessary to find a suitable operator to
describe this property. Indeed, there are various operators
suggested for this aim \cite{czachor,spin1,spin2,spin4}. We
believe that our approach potentially can be used to study the
relation between the various spin operators and the effects of LT
on the system state, provides a frame to get some predictions
about the outcomes of a Stern-Gerlach type experiment in the
relativistic situations. Here, we focused on the two spin
operators, including Czachor's and the Pauli spin operators.
Firstly, we saw that Czachor's operator is in agreement with the
Pauli operator in the limit of low velocity, both of them predict
the same outcome for a Stern-Gerlach type experiment, applied on a
spin$-\frac{1}{2}$ particle, in the lab frame. In continue, we
considered a moving frame which moves along the $x$ direction and
is connected to the lab frame by a LT \cite{hal}. It means that we
discard the relativistic effects such as the pair productions, and
in fact, one should use relativistic quantum mechanics or quantum
field theory to get a more precise results \cite{mmm,njp}.
Moreover, the lab and moving frames use the same set of
measurement directions in our setup. We found that the Pauli spin
operator predictions about the spin of particle are compatible
with the behavior of the system state in the moving frame, whiles,
Czachor's spin operator predictions differ from those of the Pauli
spin operator and the behavior of system state under LT. In
addition, we focused on the two purely bi-partite entangled
states, known as the Bell states, which include two
spin$-\frac{1}{2}$ particles moving away from each other along the
$z$ direction with the same momentum in the lab frame. Bearing LT
in mind, we evaluated the corresponding system states in the
previously mentioned moving frame. Thereinafter, we used the Pauli
operator to construct the Bell operator and the special set of
measurement directions which violates Bell's inequality in the lab
frame to its maximum violation amount ($2\sqrt{2}$). In continue,
by taking into account the same directions as the lab frame for
the Bell operator in the moving frame, we have investigated the
expectation value of the Bell operator in the moving frame. We saw
that, for a Bell state introduced in Eq.~(\ref{slab}), the
expectation value of the Bell operator in the moving frame is
decreased as a function of the boost velocity together with the
energy of particles in the lab frame. It is also found that, for
particles with low energy in the lab frame, Bell's inequality in
the moving frame is violated to the same value as the lab frame
(the maximum violation amount) in the $\beta\rightarrow1$ limit,
which is the same as the lab frame. We also addressed another Bell
state Eq.~(\ref{tlab0}) which is invariant under LT, and found out
that the expectation of the Bell operator is also invariant under
LT which is in agreement with the behavior of the system state. In
addition, the same as the one-particle system, our study shows
that the predictions of the Pauli spin operator are in line with
the behavior of system state in both of the lab and moving frames.
We have also pointed to the results obtained by considering
Czachor's operator and compared them with ours \cite{ahn}.
Finally, we found out that the Pauli spin operators are in better
agreement with the behavior of spin system in these situations.
Our study helps us make more clear the origin of differences
between the predictions about the spin behavior made by
considering Czachor's and the Pauli spin operators. It is useful
to note that although our results are in agreement with those of
three-particle non-local systems \cite{mmm}, but a Stern-Gerlach
type experiment is needed to get a decision about the quality of
validity of these results in nature.
\subsection*{Acknowledgment}
The work of H. Moradpour has been supported financially by
Research Institute for Astronomy and Astrophysics of Maragha (RIAAM) under
research project No. $1/4165-3$.

M. Bahadoran thanks the Laser Centre, Ibnu Sina, and Institute for
Scientific and Industrial Research (ISI-SIR) Universiti Teknologi
Malaysia (UTM).

\end{document}